%% file: ms.tex
\documentstyle[12pt,aasms4]{article}

\slugcomment{}
\lefthead{Reach et al.}
\righthead{Infrared Excess and Molecular Clouds}

\def\simgt{\lower.5ex\hbox{$\; \buildrel > \over \sim \;$}} 
\def\simlt{\lower.5ex\hbox{$\; \buildrel < \over \sim \;$}}

\begin{document}

\title{Infrared excess and molecular clouds:
A comparison of new surveys of far-Infrared and H~I 21-cm emission
at high galactic latitudes}

\author{William T. Reach} 
\affil{Institut d'Astrophysique Spatiale, 
 Universit\'e Paris Sud, F--91405 Orsay cedex, France \\ and \\
 Infrared Processing and Analysis Center,
 California Institute of Technology, 
 MS 100-22, Pasadena, CA 91125 }
\author{William F. Wall}
\affil{Instituto Nacional de Astrofisica, Optica, y Electonica, A. Postal 51 y 216,
Puebla, Puebla, Mexico}
\and
\author{Nils Odegard}
\affil{Raytheon STX Corporation, 4400 Forbes Blvd., Lanham MD 20706}


\begin{abstract}

We have created a map of the large-scale infrared surface brightness
in excess of that associated with the atomic interstellar
medium, using
region-by-region correlations between the far-infrared 
and 21-cm line surface brightness.
Our study updates and extends a previous attempt with the {\it Infrared
Astronomical Satellite} and Berkeley/Parkes H~I surveys. 
The far-infrared observations used here are from the
{\it Cosmic Background Explorer} Diffuse Infrared Background Experiment,
which extends far-infrared wavelength coverage to 240 $\mu$m, so that we
are reliably sampling the emission of large, thermal-equilibrium grains
that dominate the dust mass.
The H~I data are from the combined Leiden-Dwingeloo and Parkes 21-cm
line surveys.
Using the maps of excess infrared emission at 100, 140, and 240 $\mu$m, we
created an atlas and identified the coherent structures. 
These infrared excess clouds can be caused both by dust that is warmer
than average, or by dust associated with gas other than the atomic
interstellar medium. We find very few warm clouds, such as the H~II region
around the high-latitude B-type star $\alpha$ Vir.
The majority of
the infrared excess clouds are {\it colder} than the average 
atomic interstellar medium. These clouds
are peaks of column density, and their excess infrared emission is due to dust
associated with molecular gas. We identify essentially all known high-latitude
molecular clouds in the infrared excess maps, and further identify a sample of
new clouds with similar infrared properties. 
The infrared excess was correlated with CO line brightness,
allowing us to measure the ratio of H$_2$ column density to CO line integral
(i.e. the $N$(H$_2$)/$W$(CO) conversion factor)
for high-latitude clouds.
The atlas of infrared excess clouds
may be useful as a guide to regions of relatively higher interstellar column density,
which might cause high extinction toward
extragalactic objects at optical and ultraviolet wavelengths 
and confusion toward structures in the cosmic background at 
infrared and microwave frequencies.

\end{abstract}

\def\COBE{{\it COBE}}
\def\IRAS{{\it IRAS}}

\section{Introduction}

It has been known for some time that interstellar molecular gas exists away from the galactic 
midplane,
based on the presence of absorption lines of H$_2$ and other molecules in the spectra of
high-latitude stars (\cite{savage77}) and the presence of millimeter-wave emission of CO 
from regions of high optical extinction (Blitz, Magnani, \& Mundy 1984). However, the pervasiveness of
the molecular component of the local interstellar medium has not been convincingly assessed,
because of the difficulties of observing molecular gas over large areas. Absorption
line observations are restricted either to trace elements or to lines of sight toward the few 
hot, bright stars at high latitude, because the electronic ground-state transitions of both 
atomic and molecular
hydrogen are in the ultraviolet. Millimeter-wave observations of CO are restricted to 
relatively
small areas because all-sky observations are at present prohibitively expensive.
An early survey of randomly-selected lines of sight found a surface filling fraction
of 0.5\% for CO($1\rightarrow 0$) emission (Magnani, Lada, \& Blitz 1986).
A large---but, necessarily, incompletely-sampled---survey was recently performed
with a very low detection rate of 0.3\%, suggesting
that the northern galactic hemisphere is largely `devoid' of molecular gas
(Hartmann, Magnani, \& Thaddeus 1997). 
On the other hand, there remains the question of whether a survey in a
particular observable quantity, such as the brightness of collisionally-excited
rotational line emission of a particular molecule, is really
tracing all of the molecular gas. The H$_2$ and CO might have somewhat different 
spatial distributions due to their different photodissociation cross-sections and
chemistry, and the rotational energy levels of CO may not be excited in low-density 
($n({\rm H}_2)<10^3$ cm$^{-3}$) environments.

With the advent of the infrared all-sky survey by the {\it Infrared Astronomical
Satellite} (\IRAS), a new light was cast on study of the interstellar medium 
(\cite{neugebauer,low}). 
The 100 $\mu$m surface brightness was found to be well-correlated with the H~I column
density on both small and large scales (\cite{bp88}), demonstrating its value as a
tracer of interstellar gas. If the infrared emission arises from both the atomic and molecular
phases of the interstellar medium, then regions with an excess infrared emission relative
to the H~I column density are likely locations of molecular gas. This effect has been
demonstrated in Ursa Major (\cite{vht87}) and for
isolated cirrus clouds (Heiles, Reach, \& Koo 1988; 
Reach, Koo, \& Heiles 1994); in both cases
the infrared excess was found to be associated with CO emission. 
Because the infrared surface brightness and H~I column density
are known over the entire sky, it is possible to use 
the difference between the infrared map and an appropriately-scaled map of H~I
column density to produce an all-sky survey of molecular gas. This idea has been
exploited by D\'esert et al. (1988), who used the \IRAS\ 100 $\mu$m data and the
Berkeley H~I survey (\cite{hh}) to create a catalog of infrared excess clouds.

In the present paper, we study the distribution and nature of infrared excess clouds
using relatively recent data from the {\it Cosmic Background Explorer}\footnote{
The National Aeronautics and Space Administration/
Goddard Space Flight Center (NASA/GSFC) is responsible for the
design, development, and operation of the Cosmic Background
Explorer (\COBE). Scientific guidance is provided by the \COBE\
Science Working Group. GSFC is also responsible for the development
of the analysis software and for the production of the mission
data sets.} (\COBE) mission (\cite{boggess92})
and the Leiden-Dwingeloo H~I survey (\cite{hartmann97}). 
These new surveys provide higher sensitivity and higher reliability than the previous infrared
and H~I observations for large-scale emission. More importantly, the \COBE\ observations
at 100, 140 and 240 $\mu$m wavelength are a reliable measure of the emission from large,
thermal-equilibrium grains that dominate the dust mass. From a detailed study of
the infrared emission in the Orion region, it was shown that
the 100, 140, and 240 $\mu$m emission sample essentially the same dust
temperature along the line of sight (\cite{Wall}).
Our first results (\cite{reachbaas}), 
based on comparing the
\COBE\ 240 $\mu$m optical depth to the Berkeley H~I surveys, encouraged us
to pursue a more thorough comparison of the H~I and infrared data.

The large-scale distribution of molecular gas is important for a number of 
practical applications.

\noindent {\it Extinction of extragalactic objects---} 
 Extragalactic observations at visible and shorter
 wavelengths are affected by extinction even at high galactic latitude. 
 In order to estimate the extinction, which affects both the brightness and color
 of extragalactic objects, it has been
 necessary to rely on 21-cm line surveys (cf. \cite{burstein}). If an extragalactic object lies
 behind a high-latitude molecular cloud, its extinction will be significantly underestimated using
 the 21-cm line surveys. The results derived here will help observers to identify regions where
 anomalously high extinction from molecular clouds can be expected. A recent effort by
 another group (Schlegel, Finkbeiner, \& Davis 1997) 
 systematically addresses this issue,
 using the \COBE\ and \IRAS\ data to create a map of the extinction. Large-scale
 variations in the temperature and gas-to-dust ratio were calibrated by Schlegel et al.
 in a manner very similar to ours. By calculating the column density using large-scale
 average dust temperature, it is likely that the Schlegel et al. dust maps will
 underestimate the extinction toward relatively cold clouds such as the molecular clouds
 we have identified. We therefore recommend that the Schlegel et al. maps be used to
 estimate the extinction, and that our molecular cloud atlas be used as a supplement,
 warning of cold, high-extinction clouds. Lines of sight behind these clouds should
 certainly be avoided unless 1-3 magnitudes ($A_V$) of extinction can be tolerated.

\noindent {\it Shadowing of distant X-rays---} 
Interstellar clouds produce distinct shadows on the 
soft X-ray emission from the Galactic halo and the extragalactic emission (\cite{mccammon}).
These shadows have been compared to 21-cm line maps and to \IRAS\ 100 $\mu$m surface brightness
maps (Snowden, McCammon, \& Verter 1993; \cite{wang95}). In both cases, the interstellar column density is
 underestimated in the
presence of molecular clouds, because molecular gas can absorb X-rays and 
molecular clouds are relatively faint at 100 $\mu$m 
(and nearly invisible at 60 $\mu$m; \cite{Laureijs96}), compared to atomic gas.
We provide both a map of infrared excess clouds and a calibration of their 
column density in the present paper.

\noindent {\it Diffuse $\gamma$-ray emission and the N(H$_2)/W($CO$)$ factor---} 
A significant source of the $\gamma$-ray surface brightness
of the sky is due to interaction of cosmic rays with interstellar gas. The expected near-linear
proportionality of $\gamma$-rays with total column density allows a calibration of the
molecular column density, when correlated with H~I and CO maps (\cite{strong88}; \cite{digel96}). 
In the presence of H$_2$ clouds with relatively little CO emission, this calibration 
can be significantly 
biased. Here we calibrate the molecular column density using the infrared excess, and
compare the results to $\gamma$-ray studies. 
The ratio of H$_2$ column density to CO line integral is important for
assessing the vertical mass distribution of molecular gas in the
disk of our Galaxy.

\noindent {\it Relation to external galaxies---}
Low-metallicity galaxies such as the Large and Small
Magellanic clouds contain relatively fewer giant molecular clouds, and more translucent
regions---due to the lack of dust. High-latitude molecular clouds such as studied here share 
empirical
similarities with low-metallicity galaxies, and should provide insight into the nature of the 
latter. 
\section{The Correlation of Far-Infrared Emission with H~I}

\subsection{Observations}

The far-infrared data used in this work originate from the 
Diffuse Infrared Background Experiment (DIRBE, \cite{dexpsup}).
The DIRBE produced a redundantly sampled all-sky survey in 10 wavebands
from the near to far-infrared; here we concentrate 
on the 100 $\mu$m,
140 $\mu$m, and 240 $\mu$m wavebands, which are dominated by the
emission from interstellar dust. In particular, these wavebands are
dominated by the emission from the larger grains, in thermal equilibrium
with the interstellar radiation field, that dominate the mass of
dust in the interstellar medium.
Nonetheless, a significant source of large-scale emission, even
in the far-infrared, is from the zodiacal light. 
For this work, we use data from which a model for the zodiacal light
has been subtracted (\cite{reach96}; \cite{kelsall98}). To minimize detector noise,
which is significant for the 140 $\mu$m and 240 $\mu$m wavebands, we
use the average of the zodiacal-light-subtracted weekly maps over
the cold mission (44 weeks). 
The instantaneous beamsize of DIRBE is 42$^\prime$.
In order to avoid confusion between the interstellar clouds that we hope
to study and bright infrared point sources, we use data from which
all pixels containing bright and apparently unresolved structure,
as well as circular regions around the Magellanic clouds, have
been masked (\cite{arendt98}).
Furthermore, we will not consider here the portion of the sky
close to the galactic plane ($|b|<20^\circ$), where clouds from a wide range
of distances overlap on the sky.
The DIRBE data are gridded into a sky-cube projection, which we
degraded to a pixel size of $39^\prime$ so that each pixel is
nearly independent.
We assume that the point-to-point uncertainty in the infrared surface
brightness at (100, 140, 240 $\mu$m) is a quadrature combination of 
(0.15, 1.3, 0.75 MJy sr$^{-1}$) and 2\% of the surface brightness, to
account for detector noise, pixelization uncertainties, source confusion, and gain drifts. 
These uncertainty estimates were determined from inspection of the data.
At 140 and 240 $\mu$m the point-to-point uncertainty is comparable to
the expected instrument noise (cf. \cite{hauser98}), but at 100 $\mu$m
it is significantly larger, so we must have included some of the
detected small-scale sky structure of the interstellar medium.
The uncertainties are important in the present work mostly for weighting the
data, but they do scale the $\chi^2$ statistics.

The H~I 21-cm line data used in this work originate from the 
Leiden-Dwingeloo survey of the Northern sky (\cite{hartmann97})
and the Parkes survey of the Southern sky (\cite{cleary}). 
The Dwingeloo 25-m telescope has a main beam size of 36$^\prime$, which is 
well-matched to the DIRBE beamsize. 
An important aspect of the Leiden-Dwingeloo survey is the removal
of stray radiation, which is up to half of the observed signal at
high galactic latitudes. The near and far sidelobes of the telescope response
were mapped using bright point sources, then the 
sidelobe emission was estimated by convolving the sky map
with the beam (\cite{hartstray}). For this work we use the
integrated 21-cm line brightness, converted into H~I column
density assuming the emission is optically thin. This latter assumption
is unlikely to be violated at high galactic latitude unless the H~I gas is
colder than 20~K. Gas this cold must be well-shielded from the interstellar radiation field,
or else photoelectric heating will raise its temperature to at least
50~K at typical interstellar pressure. In such cold, shielded regions,
the H is rapidly converted into H$_2$, so that the H~I column density
becomes a negligible fraction of the total.
We integrated the H~I line over the range of -100 to +100 km~s$^{-1}$,
which includes the bulk of local interstellar gas but excludes high-velocity
clouds. The high-velocity clouds have been shown to be deficient 
in interstellar dust (\cite{boulwakk}); therefore they can
be ignored in the present study.
For declinations south of $-30^\circ$, we used the 21-cm survey performed
with the Parkes 64-m telescope (\cite{cleary}). 
The Parkes survey provides critical information on the southern sky,
but it does suffer from undersampling (every beamwidth in right ascension 
but every other beamwidth in declination) and incompleteness (some survey scans
missing). Furthermore, the Parkes survey has not been corrected
for stray radiation. We checked the calibration of the Parkes and Leiden-Dwingeloo surveys
by comparing their column densities where the surveys overlap;
the calibrations were in good agreement.
We assume the point-to-point uncertainty in the H~I column density is a 
quadrature sum of 
$1\times 10^{19}$~cm$^{-2}$ and 2\% of the column density, to account both for
uncertainties in the stray radiation and the gain. The uncertainties were not rigorously
determined; they are based on the estimated gain and stray radiation accuracy 
(\cite{hartmann97}).

\subsection{Region-by-region correlations}

\placefigure{fig:irhicell}

In order to identify molecular clouds in the infrared maps, we must first remove the 
infrared emission from dust associated with the atomic interstellar medium.
We presume the column density of the
atomic interstellar medium is linearly traced by the integrated
21-cm line brightness. The infrared emission, however, is not a linear
tracer of the interstellar dust because it is sensitive to the dust
temperature. We considered two methods to calibrate the dust temperature
variations: (1) calculate the column density at each pixel using the
100, 140, and 240 $\mu$m brightness ratios assuming a single dust
temperature along the line of sight, and (2) calculate the
ratio of infrared to H~I surface brightness from region to region
on the sky. The first method is simpler and, in principle, it could
detect temperature changes on angular scales as fine as 1$^\circ$. 
But in practice the 140 and 240 $\mu$m maps are very noisy and
the 100 $\mu$m map suffers from large-angular-scale zodiacal light
residuals (and all wavelengths contain some cosmic infrared background radiation).
Furthermore, we show below that the dust associated with molecular gas 
is significantly colder 
than that associated with atomic gas, so that a single dust temperature
does not apply to the lines of sight we are most interested in.
Therefore we used method (2), by dividing the
sky into `cells', with $10^\circ$ radius, on a regular grid every $10^\circ$.
Within each cell, we measured the slope and intercept of the
pixel-to-pixel infrared surface brightness as a function of H~I column density.
To illustrate the procedure, Figure~\ref{fig:irhicell} shows an example.
In the upper panel, we show the 240 $\mu$m surface brightness versus H~I column
density for the entire dynamic range of both quantities, within the region $15^\circ$
in radius centered on galactic coordinates $(l,b)=(140^\circ,35^\circ)$.
At low H~I column densities, there is a linear rise of infrared surface brightness
with H~I column densities, while at higher column densities there is a wide
dispersion. This dispersion is due to infrared emission associated with gas other
than H~I. At low column densities, interstellar H remains atomic, but
where the column density is large, the H becomes molecular.
For a uniform-density cloud, the H$_2$ column density is predicted to
exceed the H~I column density when 
\begin{equation}
N({\rm H~I}) > 2.5\times 10^{20} \left(\frac{n}{100 {\rm~cm}^{-3}}\right)^{-2}
\left(\frac{T}{80 {\rm~K}}\right)^{-1} \,\, {\rm~cm}^{-2}
\end{equation}
(\cite{rkh94}). In low-density regions, H$_2$ cannot form,
while in denser clouds such as account for the 21-cm absorption features in
front of radio sources (see review by~\cite{kulkarni88}), 
the gas readily forms molecules at column densities $>3\times 10^{20}$~cm$^{-2}$.
Figure~\ref{fig:irhicell} shows this effect: the points at higher H~I column densities
deviate systematically upward from the straight line fitted to the lower column-density points.
This upward deviation, presumably due to infrared emission from dust associated
with molecular gas, is what we will call the `infrared excess.'
We presume that on large scales, the warm and cold phases of the interstellar atomic
gas exist in pressure equilibrium, and the cold clouds have a relatively small filling
factor, so that all lines of sight cross warm regions and some (or all) cross cold regions.
In order to exclude cold clouds with column densities sufficient to form H$_2$
from our infrared-H~I correlations---which are intended to
measure the properties of the atomic medium only--- we exclude 
all points with a column density more than 
$3\times 10^{20}$~cm$^{-2}$ above the minimum column density for the `cell.'
The minimum column density is presumed to represent the warm H~I, and by allowing a
different threshold in each cell we allow for the longer path length 
through the warm gas Galaxy at low latitudes. 
The threshold column density for the Ursa Major/Ursa Minor/Camelopardalis
 region is shown as a vertical dashed line in Figure~\ref{fig:irhicell}.

\placetable{tab:irhicell}

For the low-column density pixels within a given sky `cell,' we perform a linear fit
of infrared surface brightness versus H~I column density, taking into account the
uncertainties in both quantities. For the sample cell, the points used in the fit
and their uncertainties are shown in the lower panel of Figure~\ref{fig:irhicell}.
Even at low column densities, there are some outlying positive points, 
but in no case do the fits appear to be biased by outliers.
The resulting slope and intercept for two sample cells, widely separated on the sky,
are shown in Table~\ref{tab:irhicell}. 
In both cases the fits are `good' in the sense that the reduced $\chi_\nu^2$ is essentially
unity (to within our understanding of the measurement uncertainties).
At 240 $\mu$m, the infrared-H~I slopes of the two cells are similar, but at 100 $\mu$m
the slopes are significantly different. Using the ratio of slopes at 100 and 240 $\mu$m,
(and assuming the interstellar dust emissivity varies as $\nu^2$; \cite{dl84}), 
we find that the temperature of the dust associated with H~I is $17.2\pm 0.2$~K in the
cell centered on $(140^\circ,35^\circ)$, and $19.0\pm 0.3$~K in the cell centered on
$(40^\circ,35^\circ)$. 
Even though a 2~K temperature difference between the two cells may seem unimportant,
a change in temperature from 17 to 19 K increases 
the emission at 100 $\mu$m by a factor of 2.4 and the 240 $\mu$m emissivity by a
factor of 1.7, so that the ratio of 100/240 $\mu$m emission increases by 60\%.
The ratio of dust optical depth at 100 $\mu$m (assuming $\nu^2$ emissivity) to gas column
density in the two cells are 6.4 and 4.6, respectively, in units of 
$10^{-25}$~cm$^{2}$. The absolute calibration uncertainty is $\pm15$\% in these
dust-to-gas ratios, but the uncertainty in cell-to-cell variations is
much less. Therefore, the (30\%) difference between the dust-to-gas ratios 
in these cells is statistically significant. 

\placefigure{fig:slopetv}

We have calibrated the differences in the dust temperature and dust-to-gas
ratio from region to region
by performing independent fits of the infrared surface brightness as a function of H~I column
density at each wavelength and in each sky cell. 
These fit parameters were well-constrained for most of
the sky, but there were some regions where the
infrared emission was relatively poorly-correlated with the H~I, even at low 
column densities.
If a molecular cloud complex or dust hot-spot is comparable to or larger than our cell size ($10^\circ$),
then it is not possible to calibrate the infrared-H~I slope.
These regions are generally near the galactic plane, and for the
most part they can be identified with well-known star-forming complexes. For each
of these complexes, a detailed study using higher-resolution maps is needed to
sort out the variations in gas phase and dust heating due to 
embedded stars ({\it e.g.} \cite{boulcham} for the Chameleon complex).

All-sky maps of the infrared-H~I slope and offset at 100 $\mu$m are shown in Figure~\ref{fig:slopetv}.
In cells where the fits were found to be poor, we interpolated using adjacent cells.
In cells where the fit is good, we interpolated to a finer, $37^\prime$ grid in order to 
avoid discontinuities at cell boundaries.
Two regions that stand out in the infrared-H~I slope are at $(300^\circ,+50^\circ)$ and
$(90^\circ,-40^\circ)$. The former is due to the effect of the nearby, early-type star Spica,
whose situation at high latitudes allows it to heat dust over a very large 
apparent area (\cite{reynoldsspica}), 
while the latter is due to the MBM 53/54/55 complex of molecular clouds (\cite{mbm85}). 
Our cell size was chosen so that regions like these would not completely
disappear once the infrared emission associated with H~I is subtracted. 
In addition to these relatively compact regions, 
there are real variations in the 100 micron-HI slope
 due to temperature variations from place to place of 
 the interstellar dust associated with atomic gas.
These variations appear in Fig.~\ref{fig:slopetv} as large-scale structures. 
One noticeable problem is at southern declinations, where the H~I data from the Parkes
survey was used. The infrared emission per H atom is lower there than average, 
which could be due in part to 
stray radiation effects in the Parkes survey.

\subsection{Implications for the far-infrared background}

\placefigure{fig:slopehis}
\placefigure{fig:slopehisb}

For each of the far-infrared wavelengths we have considered here, we find that a 
positive residual sky brightness remains after subtracting the infrared emission correlated with
the H~I column densities. 
Histograms of all the zero-intercepts of the infrared-H~I correlation for out grid of sky cells
are shown in Fig.~\ref{fig:slopehisb};
it is clear that the zero-intercept is consistently positive.
This same result has also been found from the correlations
of infrared brightness with H~I column density in the Lockman Hole and around the North Ecliptic Pole
(\cite{arendt98}), and for the entire low-column-density portion of the northern sky
(\cite{boulanger96}). 
At least part of the residual emission is related to the Solar System, because one can
see the distinctive pattern of the ecliptic plane in Fig.~\ref{fig:slopetv}. 
However, the amplitude of the residual
zodiacal light, based on its ecliptic latitude dependence, is small. Furthermore, the
spectrum is inconsistent with the zodiacal light, being relatively bright at 240 $\mu$m,
which argues against a Solar System origin (\cite{Dwek98}).
The far-infrared emission remaining after the subtraction of the emission from interstellar dust
could be the extragalactic background due to unresolved galaxies,
with obvious cosmological significance; alternatively, it could be due to a component of
our galaxy with infrared emission but little atomic gas.
We will address the cosmological issue only briefly here, as it is the primary focus of other recent papers
(\cite{puget96}; \cite{hauser98}). 
In the present work, we have allowed for the possibility that the infrared
emission per unit H~I column density varies from place to place.
In the presence of uncertainties in both variables, the slope and zero-intercept of a correlation
tend to be correlated. Because the H~I column density is always positive,
the slope and intercept are negatively correlated; furthermore,
the intercept is biased toward positive values where the correlation is weak.
Comparing the slope and offset of all of the sky cells with a significant fit,
we find a near-perfect anti-correlation.
This could potentially lead to a false detection of the cosmological background, especially if,
at low column densities, the interstellar dust is associated with an interstellar medium other
than atomic gas.

\placetable{tab:slopehis}

The infrared emission per unit H~I column density has a well-defined mean value,
with few regions of abnormally high or low values. Table~\ref{tab:slopehis} lists
the mean and rms dispersion of the infrared emission per unit H~I column density 
and the zero-intercept of the infrared-H~I correlation at
100, 140, and 240 $\mu$m. It is clear that the range of slopes and intercepts obtained over
the high-latitude sky is significantly larger than the uncertainty of an individual value, so 
that the range of slopes is not due to random measurement or fitting uncertainties.
The wavelength-dependence of the slopes in the two sample regions (Tab.~\ref{tab:irhicell})
indicates that this is mostly a dust temperature effect.
This is a gradual, large-scale effect, that is not confined to the few outstanding cells.
The cells with an infrared-H~I slope more
than 1.5 times the rms away from the mean are shaded
in Fig.~\ref{fig:slopehis} and Fig.~\ref{fig:slopehisb}.
It is evident that the outlying regions in terms of infrared-H~I slope are also outlying
in terms of the zero-intercept (as expected because of the anti-correlation of slope and
intercept mentioned above).
Over the entire sky, there is a significant, positive sky brightness that cannot be explained by
interstellar dust associated with H~I.
If the range of slopes shown in Fig.~\ref{fig:slopehis} is a {\it real} effect due to the
interstellar medium, then the range of offsets in Fig.~\ref{fig:slopehisb} is also due
to the Galaxy.
The inferred brightness of an isotropic extragalactic background must then be lower 
than the {\it lowest} offset value. In principle, variations in the infrared-H~I correlation
on scales smaller than those chosen for the present work could lead to an even
wider range of slopes and offsets. 
The {\it dispersion} of the putative infrared background brightnesses, listed in the last
column of Table~\ref{tab:slopehis} precludes more than a `2-$\sigma$' (loosely speaking)
confidence level detection of the infrared background, unless more restrictive assumptions about the
galactic emission are made. The significance of this effect is discussed in
more detail by Hauser et al. (1998) and Shafer et al. (1998). Fortunately, the separation
of the galactic and cosmological contributions to the far-infrared sky brightness can
be resolved using higher-resolution and sensitivity far-infrared observations, 
such as have recently been performed with the {\it Infrared Space Observatory}
(\cite{kawara,pugfirback})
and will eventually become possible with the NASA Space Infrared Telescope Facility and
the ESA {\it Planck Surveyor}.
These observations should reveal whether the brightness of the faintest parts of the sky is produced by
an ensemble of extragalactic sources or by interstellar cirrus. 

\section{The Distribution of Infrared Excess}

\subsection{Definition of infrared excess}

The infrared excess is defined to be the observed infrared surface brightness minus
the contributions from dust associated with the atomic interstellar medium, the
zodiacal light, and the cosmic infrared background. We calculate the infrared
excess using
\begin{equation}
I_{\nu}^{ex} = I_{\nu} - S * N({\rm H~I}) - O,
\end{equation}
where $S$ and $O$ are the locally-interpolated slope and offset of the infrared-H~I correlation.
The offset $O$ contains residual zodiacal light and the cosmic infrared background,
and is not useful for measuring properties of the interstellar medium.
We created infrared excess maps independently for the 100, 140, and 240 $\mu$m wavebands,
so that we can measure the far-infrared colors of the infrared excess.
The infrared excess map at 240 $\mu$m is shown in an all-sky projection in Figure~\ref{fig:extv}.
The infrared excess is `patchy,' with significant structure on scales smaller than the
size of the `cells' within which the infrared-H~I slopes were determined. We have checked
that the maps are not very different if the cell size is twice as large as the one we finally
adopted. 
The significance of a given value of the infrared excess can be determined statistically
from its probability distribution. Figure~\ref{fig:exhis} superposes the histograms of
the positive and negative pixels in the infrared excess map. It is clear that the probability
distribution is not random, and that the {\it positive} values significantly outnumber the
negative values when $I_{100}^{ex} > 0.3$~MJy~sr$^{-1}$, which we will use as the
lower limit to define infrared excess clouds. 

\placefigure{fig:extv}
\placefigure{fig:exhis}

\subsection{Atlas of infrared excess clouds}

A table of clouds would not suffice to describe the infrared excess, because the emission 
has some continuity over large scales.
High-latitude molecular clouds have been shown to be associated with large
H~I features (Gir, Blitz, \& Magnani 1994), suggesting their formation is related to events or processes
(like supernovae and stellar winds) that organize the interstellar medium on large scales.
For presentation here, and practical utility to other researchers, we present the infrared
excess as an atlas consisting of 8 maps with simple coordinate projections.
We have preferred a Cartesian projection for latitudes $20^\circ < |b| < 50^\circ$, so that positions can be
easily located with a ruler (Fig.~\ref{fig:exmaps}). We also made maps of the two galactic polar caps,
in a simple orthographic projection (Fig.~\ref{fig:exmappole}). The maps we present here
are for 100 $\mu$m wavelength, the highest-sensitivity of the DIRBE far-infrared
wavebands. The maps at 140 and 240 $\mu$m are generally similar.
At 100 $\mu$m, there are more infrared excess clouds, because both {\it warmer-than-average} 
and {\it molecular} clouds produce infrared excess at 100 $\mu$m. However, the vast majority of 
structures in the far-infrared excess maps are shown in the next sections
to be {\it cold}.

\placefigure{fig:exmaps}
\placefigure{fig:exmappole}

In order to study the properties of the `clouds' and regions of sky with 
infrared excess, we have identified two lists of positions and apertures within
which to perform photometry. The first list is the catalog of previously-known
molecular clouds compiled by Magnani, Hartmann, \& Speck (1997). (We combined entries 
that would be redundant at our low resolution.) These are indicated on the
individual panels of the atlas. Second, we made a list of some of the remaining peaks
in the infrared excess maps, with brightness comparable to the known molecular clouds.
This list was created by hand, upon inspection of both 100 $\mu$m and 240 $\mu$m
infrared excess maps, and it is incomplete.
The `new' clouds were named in the format DIR$lll\pm bb$ where $lll$ is the galactic
longitude and $bb$ is the galactic latitude; the prefix `DIR' stands for `diffuse
infrared' as in the name DIRBE.
Third, for reference, we labeled many peaks in the infrared excess maps
due to bright point sources, most of which are nearby galaxies (Odenwald, Newmark,
\& Smoot 1997).
The list of previously-known clouds is shown in Table~\ref{tab:excatknown},
and the newly-identified clouds are shown in Table~\ref{tab:excatnew}. 
For each cloud, we averaged the infrared excess surface brightness at 100 and 240 $\mu$m
within the cloud boundary. 
Only clouds detectable both at 100 $\mu$m and 240 $\mu$m are listed.
The total infrared excess flux at 100 $\mu$m,
the color temperature obtained from the 100/240 $\mu$m ratio (assuming emissivity proportional to
$\nu^2$), and the angular diameter of each cloud is 
shown in Tables~\ref{tab:excatknown} and~\ref{tab:excatnew}.

\placetable{tab:excatknown}
\placetable{tab:excatnew}


The infrared excess clouds in general are relatively bright at 240 $\mu$m, and
their far-infrared color temperature is significantly lower than that of average
atomic clouds. This is illustrated in Figure~\ref{fig:extemphis}, where histograms
of the color temperatures of infrared clouds are shown. 
The fact that the vast majority of infrared excess clouds are {\it cold}
confirms that the infrared
excess is generally {\it not} due to clouds that are warmer than average.
Indeed, because the clouds are colder than average, they tend to be
under-luminous in infrared emission. This effect has also been found by
Lagache et al. (1997), who showed that molecular regions at high latitude
can be identified by (1) deficiency in 60 $\mu$m emission
and (2) low 140/240 $\mu$m color temperatures relative to atomic cirrus. 
One of the main uncertainties that has been faced by past studies of
infrared excess as a tracer of molecular gas ({\it e.g.} \cite{dbb88}; \cite{rkh94}) 
was the uncertainty as to whether the infrared surface brightness at 60 $\mu$m, 
observed by \IRAS\ and often compared to the 100 $\mu$m brightness, 
traces variations in column density or
dust temperature. Using the \COBE\ observations at 240 $\mu$m, this
problem has now been resolved.

\subsection{Warm infrared excess clouds}

While they are rare, there are some infrared excess clouds that are {\it warmer}
than average. The single most
prominent example is the large H~II region around
the nearby B star $\alpha$ Vir, or Spica, which was first discovered by its extensive
H$\alpha$ emission (\cite{reynoldsspica}). The Spica H~II
region is exceptional, in that it is produced by one of the closest---at $d=80\pm 5$ pc 
(\cite{esa97})---early-type stars; it is at high galactic latitude, where it
can be easily separated from other sources; and the star is presently in a region of
relatively low interstellar medium density, so that its H~II region covers a region
some $15^\circ$ across. The dust in the Spica H~II region is significantly
warmer than typical interstellar cirrus, and it is unusual in that it also has 
12 and 25 $\mu$m emission significantly higher than expected from the
100--240 $\mu$m emission, suggesting a modified grain size distribution (\cite{zagury}).
There are no other `warm' infrared excess clouds comparable to the Spica H~II region
in size and brightness, most likely because no other early-type star is as
advantageously situated.

One significant `warm' infrared excess cloud that has not been noted before is DIR015+54. 
This is one of the brightest 100 $\mu$m infrared excess clouds at high positive galactic latitudes
(Fig.~\ref{fig:exmappole}a), but it is not detected at 240 $\mu$m (and therefore is
absent from Table~\ref{tab:excatnew}).
In order to see
the structure of this cloud at higher angular resolution, we extracted {\it IRAS}
images at 12, 25, 60, and 100 $\mu$m from the {\it IRAS Sky Survey Atlas} ({\it ISSA}).
The cloud is relatively bright at 60 $\mu$m, with a local-background-subtracted
surface brightness, averaged
over the central 30$^\prime$, of 0.8 MJy sr$^{-1}$. The ratio of
60/100 $\mu$m brightnesses is 0.46, a factor of 2 higher than that of
atomic `cirrus' and much higher than that of molecular clouds. 
The {\it ISSA} 60 $\mu$m image is shown in Figure~\ref{fig:myst}. 
The cloud is structured  in a form qualitatively suggestive of a bow-shock
produced by an object moving toward the northeast. Bow-shocks were
discovered in the {\it IRAS} data due to the stand-off between the interstellar
medium and the stellar winds of early-type stars with significant transverse
velocities (\cite{dvb88}), and their morphologies and 60/100 $\mu$m colors
are comparable to the cloud shown in Figure~\ref{fig:myst}. However, there
is no early-type star in the vicinity that could plausibly create this nebula,
either as an H~II region or a bow shock. Such a star
would be very bright at visible and ultraviolet wavelengths, owing to the
low extinction along the line of sight and the proximity needed to heat 
a region of such a large angular size. 
There is a peak of 60 $\mu$m
emission approximately at the location where a wind-source would be located.
The nature of this emission peak, listed in the {\it IRAS Faint Source Catalog} as
F15101+1206 (\cite{fsc}), is presently unknown, though statistically it is 
likely to be extragalactic (Reach, Heiles, \& Koo 1993).
Apart from some of the peaks in the nebula, which appear in the 
{\it IRAS Faint Source Catalog}, nearby 
there are only some galaxies and stars of type A and later
in the SIMBAD database.
We find that there is a faint optical counterpart of F15101+1206 in the
{\it Palomar Observatory Sky Survey}\footnote{Based on photographic data of the National Geographic Society -- Palomar
        Observatory Sky Survey (NGS-POSS) obtained using the Oschin Telescope on
        Palomar Mountain.  The NGS-POSS was funded by a grant from the National
        Geographic Society to the California Institute of Technology.  The
        plates were processed into the present compressed digital form with
        their permission.  The Digitized Sky Survey was produced at the Space
        Telescope Science Institute under US Government grant NAG W-2166.
}, consistent with the idea that F15101+1206 is extragalactic.
There is no optical counterpart for the diffuse emission of DIR015+54.
The infrared colors of DIR015+54 are similar to those of the nearby galaxy NGC 6822
(\cite{israel96}), suggesting perhaps that the cloud could in fact be another low surface
brightness member of the Local Group.
We inspected the individual H~I spectra from the Leiden-Dwingeloo 21-cm line survey near
this position, and we find that the cloud is clearly present, with a `normal' LSR velocity of 
-17 km~s$^{-1}$ and both narrow and wide line components (dispersions 3 and 15 km~s$^{-1}$,
respectively). We suspect therefore that DIR015+54 is a Galactic cloud that happens to
be somewhat brighter than average in the far-infrared, and much brighter than
average at 60 $\mu$m.
The nature of this cloud and the other large, `warm' clouds in the {\it COBE}/DIRBE and {\it IRAS}
data remains mysterious. They are either locally-heated regions or regions
with unusual grain size distributions, but we do not know why;
this should be investigated further. 

\subsection{Comparison with known molecular clouds}

Our atlas of infrared excess clouds reveals essentially {\it all} previously known
high-latitude molecular clouds in the literature (\cite{magnani97}). 
First, consider the clouds originally found by Magnani, Blitz, \& Mundy 
(1985, hereafter MBM);
these clouds were first identified as regions of extinction on the Palomar
Observatory Sky Survey plates, and then they were detected in CO($J=1\rightarrow 0$)
emission with radio telescopes.
At our low angular resolution, the MBM clouds comprise essentially 36 individual regions.
Of these, well-formed clouds with significant infrared excess ($I^{ex}_{100}>1$ MJy sr$^{-1}$) were clearly detected
from 34 regions, for a 94\% detection rate. MBM 19 and MBM 9, the two regions for which we
do not see an infrared `cloud', both are in regions of significant infrared excess, but there is
no well-defined cloud at their position. Indeed this is somewhat true of some of the other 
clouds as well:
the center of the infrared excess region does not coincide with the center of the MBM cloud.
We suspect that the MBM positions are relatively dense cores (with well-defined extinction
on the optical plates), while the overall cloud structure is better represented by the infrared excess map.

\placefigure{fig:extemphis}

Our atlas of infrared excess is significantly different from the catalog of 515
infrared excess clouds obtained by D\'esert et al. (1988) [DBB]. We have attempted to
associate each entry of Tables~\ref{tab:excatknown} and~\ref{tab:excatnew}
with a cloud in the DBB catalog. 
For 62 \% of the MBM clouds, we find corresponding entries in the DBB catalog.
This is somewhat
higher than the 47\% detection rate determined by DBB, because of the way we grouped MBM clouds
into complexes and considered a match if the DBB and MBM positions fall within the same cloud 
in Figs.~\ref{fig:exmaps} or~\ref{fig:exmappole}. Specifically, this means we consider the MBM clouds to be
much larger than the high-extinction and CO-emitting cores.
The higher detection
rate of MBM clouds in our atlas is due to
the higher quality input data that we used; both
infrared and HI are of substantially higher quality
than the data available to DBB.  
Many of the clouds that are not listed in DBB are very bright in our atlas, including both well-known
(MBM) and new clouds. On the other hand, a significant fraction of DBB clouds are not confirmed by
the present analysis. In total, we find counterparts for 45\% of the DBB clouds and hints for another 7\%,
leaving 48\% with no evident infrared excess in our maps.
Of the unconfirmed clouds, 58\% have a `significance' criterion less than 5 listed in the DBB table, but
10\% (mostly H~I artifacts) are listed with `significance' greater than 7. 
The uncertainties in the
infrared excess method---especially with the lower-quality infrared and H~I data used by DBB, but
also with the results presented in this paper---are dominated by systematic uncertainties rather than 
random noise. We have therefore chosen to make a relatively short table of 81 selected clouds, 
and to present the sky maps for use by others.

\placefigure{fig:exheith}

The interstellar medium is not an ensemble of clouds with well-defined boundaries---it is
coherent over large angular scales. Therefore a comparison of cloud catalogs can be misleading.
There are few regions where the CO($J=1\rightarrow 0$) line has been mapped in emission on scales
large enough to make a meaningful comparison with our atlas. The largest-scale high-latitude CO
map of which we are aware was made with a 1.2-m telescope, whose large beam (9$^\prime$)
is well-suited to comparison with our results. 
A large portion of the Ursa Minor-Ursa Major region has been covered in a number of surveys, 
including one very large, regularly sampled grid (\cite{heith93}). We have projected our infrared
excess map onto the grid of the CO map assembled by Heithausen et al. (1993) for detailed
comparison---see Figure~\ref{fig:exheith}. The agreement between the 
contour of integrated CO line brightness, $W({\rm CO})$,
and the infrared excess contour is remarkably good, and the infrared excess map  contains
no significant clouds in regions without CO. 
Figure~\ref{fig:excorr} shows a pixel-by-pixel scatter diagram for these maps.
A linear fit, taking into account the statistical uncertainties of 0.5 K km s$^{-1}$
in $W({\rm CO})$ and 0.3 MJy sr$^{-1}$ in $I_{100}^{ex}$ yields the following:
\begin{equation}
W({\rm CO}) = \left(-0.19\pm 0.05\right) + (1.17\pm 0.05) \left(I_{100}^{ex} /  {\rm MJy~sr}^{-1}\right) \,\,{\rm K~km~s}^{-1},
\end{equation}
where only statistical uncertainties were accounted for.
The correlation of 100 $\mu$m emission with H~I column density in this region 
(which largely overlaps with that shown in Fig.~\ref{fig:irhicell}) yields
\begin{equation}
I_{100} = 0.47\pm 0.1 + \left(0.62\pm 0.03\right) \left( N({\rm H~I}) / 10^{20} {\rm~cm}^{-2}\right) \,\, {\rm MJy~sr}^{-1}.
\end{equation}
If the infrared emission per H nucleus were the same in atomic and molecular gas, then we find that the
H$_2$ column density per unit integrated CO($1\rightarrow 0$) line brightness is
\begin{equation}
N({\rm H}_2) / W({\rm CO}) = \left(0.7\pm 0.1\right) \times 10^{20} \,\, 
{\rm cm}^{-2}\, {\rm K}^{-1}\,{\rm km}^{-1}\,{\rm s}.
\end{equation}

If they were sufficiently orthogonal, we could determine the separate emissivities of the atomic
and molecular gas using a multiple linear regression between 
the infrared, H~I, and CO surface maps over the parts of the region that have
been observed at all three wavelengths.
Table~\ref{tab:excorr} shows the slopes from fits of the form
\begin{equation}
I_\nu = a N({\rm H~I}) + b W({\rm CO}) + c
\end{equation}
for several infrared wavelengths.
It is evident that $a$ and $b$ depend on wavelength differently, and the interpretation is straightforward:
the dust mixed with the atomic gas has a different temperature from the dust mixed with the
molecular gas. The color temperature inferred for the dust mixed with the atomic gas is
$T_1=17.0\pm 0.3$~K, and for the dust mixed with the molecular gas it is $T_2=14.7\pm 0.2$ K.
The observation that atomic and molecular clouds have different infrared spectra implies
that their relative column densities
cannot be simply determined using observations at a single wavelength. 
This has an important practical implication. Na\"{\i}vely, the ratio 
\begin{equation}
X_{naive}=0.5 b/a
\end{equation}
gives the column density
of H$_2$ per unit CO line integral; values of $X_{naive}$ are listed in the last column of
Table~\ref{tab:excorr}. It is evident that the na\"{\i}vely inferred value of $X$ changes by a factor of 2 depending on
the wavelength, which is inconsistent with the idea that the infrared emission is a linear tracer of
the total column density.
We can, however, calibrate this effect using the results obtained here. We assume that within 
the molecular phase, the gas is completely molecular and the dust-to-gas ratio is the same
as in the atomic phase. 
Then the  column density of H$_2$ per unit CO line integral is
\begin{equation}
X = \frac{1}{2} \frac{b}{a} \frac{B_\nu(T_1)}{B_\nu(T_2)},
\end{equation}
(\cite{magnani95})
which results in the last column of Table~\ref{tab:excorr}. When calibrated this way,
a self-consistent description of the infrared emission from the atomic and molecular phases
could be obtained from observations at a single wavelength---if the temperature contrast between 
the atomic and molecular components were the same everywhere.  
Clearly, observations at 240 $\mu$m 
are the best infrared tracer of molecular gas, because the 240 $\mu$m emission is less
sensitive to temperature changes.

The H$_2$ column density per unit CO line integrated intensity has been measured by several techniques
and for several different regions, with a fairly wide range of results. 
The `canonical' method is to correlate $\gamma$-ray emission---due to interaction
of cosmic rays with nuclei---with the CO and H~I emission simultaneously.
This correlation has been performed in the galactic plane, where the CO emission is
dominated by giant molecular clouds, 
and  $X=(2.3\pm 0.3)$ has been found
\footnote{Units for $X$ throughout this paper are $10^{20}$ cm$^{-2}$ [K~km~s$^{-1}$]$^{-1}$.}
(\cite{strong88}). The correlation of $\gamma$-ray emission and CO line brightness
in the Ursa Major/Ursa Minor region yielded $X=(0.92\pm 0.1)$ (\cite{digel96}).
The $\gamma$-ray method, like the `na\"{\i}ve' infrared method, assumes that the
$\gamma$-ray emission properties of the atomic and molecular clouds are the same.
In fact, recent observations of lower-energy $\gamma$-rays indicate that the value
of $X$ inferred depends on the energy of the $\gamma$-rays, with values ranging
from (1.1--4) (\cite{strong94}). 
If the difference is due to processes that only operate at low energies, then the 
`canonical' value of 2.3 holds, for it was determined at high energies; however,
until this is understood, there remains the possibility that even the `canonical' $X$ 
is uncertain by a factor of 2.

The infrared method of locating and estimating the column density of H$_2$ is
especially useful because a wide range of physically different types of cloud
can be observed. 
For the galactic center, Sodroski et al. (1995) found that the value of $X$ 
is an order of magnitude lower than in the outer Galaxy.
Their analysis takes into account the higher dust temperature in the galactic
center, because they used the dust optical depth from 140 and 240 $\mu$m observations.
At high latitudes, the molecular clouds are smaller, so one might anticipate a difference
in the H$_2$ column density per unit CO line integrated intensity due to the higher radiation
fields that can penetrate a smaller cloud.
An analysis very similar to our multiple linear regression was performed for the Ursa
Major clouds using higher-angular-resolution 21-cm line observations with the Effelsberg 100-m
telescope and {\it IRAS} 100 $\mu$m data (\cite{vht87}).
Their infrared map clearly revealed the presence of the clouds that were not evident in
the 21-cm line map, and the combined
H~I and CO maps were found to reproduce the infrared brightness, with 
$X_{naive}=(0.5\pm0.3)$. Using the temperature of atomic and molecular
clouds we found for the entire North Celestial Pole region, 
\begin{equation}
X=3.8 X_{naive}=(1.9\pm1.1) \times 10^{20} \, {\rm cm}^{-2}\,[{\rm K~km~s}^{-1}]^{-1}. 
\end{equation}
The corrected $X$ for the Ursa Major clouds
is consistent with the `canonically' adopted $\gamma$-ray results.

For some clouds, however, the H$_2$ column density per unit CO line integral is 
significantly different from canonical values. The cloud HRK236+39 was observed
in the 21-cm line with the Arecibo 305-m telescope, which provides the highest-resolution
single-dish observations of an isolated, diffuse interstellar cloud (\cite{rkh94}).
For this cloud, the linear regression of the H~I, CO, and infrared maps 
resulted in the `na\"{\i}vely' determined value $X_{naive}=(0.16\pm0.03)$. 
This low value is not due to a temperature effect: we find in the present work
that the color temperature of the far-infrared excess emission is 16.5~K, only
slightly colder than that of atomic gas. In order to explain the low $X_{naive}$ with
a temperature difference alone, the temperature of the molecular gas would have
to be 13.2~K. While this might seem similar to the observed 16.5~K, it is in fact
clearly ruled out: for the same 100 $\mu$m brightness,
the 240 $\mu$m emission of the cloud would be more than 4 times
brighter than observed.
For HRK236+39, comparison of the H~I, CO, and infrared maps clearly shows that
there is a component of the infrared emission that is not traced by H~I 21-cm or
CO($1\rightarrow 0$) line emission. 
The infrared excess exists in the central $\sim 45^\prime$ of the cloud, while
the CO is concentrated even more to the center, in clumps with size $\sim 2^\prime$.
Similar results have been found from a comparison of infrared and CO maps of the Chameleon 
region by Boulanger et al. (1997). They found that the large clouds, with extensive CO 
emission, $X_{naive}=$(0.6--1.0), which could easily be
explained by a somewhat lower dust temperature in the molecular clouds. The large scatter
in the comparison of CO and extinction maps---which are independent of the dust
temperature---was found to be due to spatially coherent regions where the gas
is primarily H$_2$ but the CO abundance is low.

\placetable{tab:excorr}

Any map of molecular gas, even over a very restricted region,
would allow us to calibrate the infrared excess. Where it has been observed,
the extinction of background starlight provides a temperature-independent 
tracer of the total column density. Comparison of extinction and CO maps
for high-latitude clouds yields $X=$(0.4--6.4) (\cite{heith90}; 
Magnani, Blitz, \& Wouterloot 1988),
but it is unclear whether the large
scatter is due to real cloud-to-cloud variations or calibration difficulties in
the extinction maps (which are obviously uncertain when the extinction is
appreciably more than unity).
The H$_2$ molecule is symmetric, so that its rotational transitions must be
quadrupolar,
and the lowest-energy transition has an energy of 500~K, which cannot be excited
under typical interstellar conditions. Absorption line studies are able to trace cold, diffuse gas,
but the paucity of suitable background stars makes comparison to large-beam observations
problematic. A promising method is to observe the low-energy radio transitions of the
CH molecule, because CH forms readily---but only where the H is molecular.
Magnani \& Onello (1995) have observed a set of both translucent and dark
molecular clouds in the 9-cm lines of CH, and used the empirical correlation
of CH line brightness with extinction to obtain the H$_2$ column density. They
find that $X=$(0.3--6.8) for translucent high-latitude clouds: a substantial range,
as was found from the extinction studies. 
The value of $X$ was also found to vary within clouds, where it was observed for
more than one position. 
A detailed comparison of our results with those of 
Magnani \& Onello (1995) is complicated by the fact that the clouds are structured on
scales smaller than our beam, while the CH lines have not been mapped over the
clouds.
While there is clearly an infrared excess counterpart of each of the clouds
in their sample, only 11 lines of sight are suitable for comparison; we find that
the ratio of 240 $\mu$m infrared excess to H$_2$ column density (inferred from
the CH line brightness) ranges from 0.4 to 2 MJy~sr$^{-1}$/($10^{20} $~cm$^{-2}$).
It is unclear whether the range of values is due to beam size effects or a
true variation from cloud to cloud, but this comparison at least affords a
calibration of the infrared excess independent of CO. For comparison, the
calibration of 240 $\mu$m emission per H$_2$ for the North Celestial Pole
region yields 
\begin{equation}
I_\nu(240\,\mu{\rm m})/N({\rm H}_2)=b/X=1.0\pm0.2\,\, 
{\rm MJy~sr}^{-1}\,{\rm (10}^{20}\,{\rm cm}^{-2}{\rm)}^{-1}
\end{equation}
using Table~\ref{tab:excorr}, 
which is within the range given by the CH calibration.

\placefigure{fig:excorr}

\section{Conclusions}

The infrared excess clouds, which were found after removing emission associated with 
the atomic interstellar medium, are predominantly molecular clouds. 
The dust in these clouds is colder than dust in diffuse atomic clouds, for the same
reason that the gas is molecular: the outer layers of the cloud shield the center from the
dust-heating and molecule-dissociating effects of the interstellar radiation field.
These molecular clouds fill only a small fraction of the high-latitude sky. 
The column density of the infrared excess clouds can be estimated using the calibrations
discussed above, which may be rewritten as 
\begin{equation}
I_{100}^{ex}/N({\rm H}_2) = (0.26\pm0.05) \times {\rm 10}^{-20}\,{\rm cm}^{2}
{\rm MJy}\,{\rm sr}^{-1}.
\end{equation}
We can estimate the mass surface density of the infrared excess clouds, using
the difference between positive and negative excess for $|b|>20^\circ$
(Fig.~\ref{fig:exhis}). 
The surface density is about 0.3~$M_\odot$~pc$^{-2}$
(including 40\% He by mass), roughly equally partitioned among regions of low and
high brightness.
(We adjusted for the part of the sky at $|b|<20^\circ$ by assuming it has the same
average infrared excess as the rest of the sky.)
The mass density that we find is
an order of magnitude larger than that found in an unbiased, wide-field CO survey (\cite{hartmag})
of the northern sky, but it is comparable to a previous estimate---also based on CO---of the total 
mass of the known molecular clouds at high galactic latitude (\cite{magnani97}). We suspect that
the unbiased survey missed much of the molecular gas because either there was no CO
mixed with the diffuse H$_2$ or the CO rotational levels were subthermally excited 
(because of low H$_2$ volume density). We do {\it not} find a great disparity between the northern
and southern galactic hemispheres (cf. \cite{hartmag}); according to the infrared excess, the southern
galactic hemisphere has about 30\% more molecular gas than the northern hemisphere. 
The high-latitude infrared-excess
clouds comprise a significant fraction of the total mass of molecular gas in the solar neighborhood.

The infrared excess maps can be used to provide a guide to regions of anomalously high
extinction. The extinction calculated from the H~I column density
(cf. \cite{burstein}) can significantly underestimate the total extinction in these regions.
If we assume that the dust in the infrared-excess clouds has
the same extinction cross-section per H-nucleus as diffuse clouds for which this
quantity has been measured (Bohlin, Savage, \& Drake 1978), then the extinction due to infrared excess clouds
can be obtained from Fig.~\ref{fig:exmaps}, after multiplying the
100 $\mu$m surface brightness by  $A_V/I_{100}^{ex}\sim 0.027$~mag~MJy$^{-1}$~sr.

\acknowledgements
WTR was supported by NASA's \COBE\ project during the early part of this
work, through a contract to the Universities Space Research Association.
In searching for counterparts for the objects found in this work, we use the {\it SkyView} facility
at NASA Goddard Space Flight Center; {\it SkyView} is supported by NASA ADP grant NAS 5-32068.
We also made use of the SIMBAD database, which is maintained by the Centre de Donnees
astronomiques de Strasbourg (CDS), France.
We thank Xavier D\'esert, Fran\c{c}ois Boulanger, Rick Arendt, Janet Weiland, Tom Sodroski,
and Mike Hauser for their comments and suggestions.

\clearpage

\figcaption[irhicell.ps]{Infrared surface brightness at 240 $\mu$m observed by 
DIRBE plotted versus the atomic gas
column density from the Leiden-Dwingeloo 21-cm line survey for the Ursa Major/Ursa Minor/Camelopardalis
region. In the top panel, nearly the full dynamic range of H~I and infrared brightness is shown;
56 points, out of 2393 total, with $N({\rm H~I})>1.6\times 10^{21}$ cm$^{-2}$ were excluded.
Error bars are shown for a few points for illustration.
The vertical dashed line indicates the threshold column density below which a linear fit was
performed.
The bottom panel shows the same data, but with restricted dynamic range covering only the
points used in the fit, $N({\rm H~I})<3.75 \times 10^{20}$ cm$^{-2}$. 
The linear fit to the low-column density points is shown  in both panels as a diagonal solid line.
In this paper we concentrate on the positive residuals from the linear fits, which we attribute
to dust associated with ionized and molecular gas.
\label{fig:irhicell}}

\figcaption[slopetv.ps]{All-sky maps of the slope 
{\it (a)} and offset {\it (b)} of the correlation between
100 $\mu$m surface brightness and H~I column density. The maps are shown in a
galactic-coordinate Mollweide projection with the galactic center
at the center. Meridians are drawn every $60^\circ$, and parallels are drawn 
every $30^\circ$.
The greyscale ranges are labeled to the right of the images;
for panel {\it (a)} in
MJy sr$^{-1}$ / (10$^{20}$ cm$^{-2}$), and for panel {\it (b)} in
MJy sr$^{-1}$.
\label{fig:slopetv}}

\figcaption[slopehis.ps]{Histograms of the 
slopes of the infrared-H~I correlations for sky patches
with 10$^\circ$ radius.
Each panel shows the results from a different DIRBE waveband:
{\it (a)} 100 $\mu$m,
{\it (b)} 140 $\mu$m,
{\it (c)} 240 $\mu$m.
The `outlying' values of the slope, defined as being more than 1.5 times the rms away 
from the mean, are shaded.
\label{fig:slopehis}}

\figcaption[slopehisb.ps]{Similar to previous figure, 
but this time for the {\it offsets} of the
infrared-H~I correlations for sky patches with 10$^\circ$ radius.
Each panel shows the results from a different DIRBE waveband:
{\it (a)} 100 $\mu$m,
{\it (b)} 140 $\mu$m,
{\it (c)} 240 $\mu$m.
The sky patches with `outlying' slope are shaded.\label{fig:slopehisb}}

\figcaption[extv.ps]{All-sky map of the 
residual 240 $\mu$m surface brightness after subtracting the
emission correlated with the H~I column density. 
The map is shown in a galactic-coordinate Mollweide projection as Figure 2; 
meridians are drawn every $60^\circ$ and parallels every $30^\circ$.
The greyscale range is illustrated by the bar, with units MJy~sr$^{-1}$, 
to the right of the map.
\label{fig:extv}}

\figcaption[exhis.ps]{Histograms of the infrared
excess at 100 $\mu$m. The upper histogram shows the distribution of the positive values of
infrared excess, and the curve beneath it shows the distribution of the negative values.
The positive values clearly dominate above an
infrared excess of 1 MJy~sr$^{-1}$. \label{fig:exhis}}

\figcaption[exmaps.ps]{An atlas of the 100 $\mu$m infrared 
excess in galactic-coordinate Cartesian projections,
for absolute galactic latitudes 20$^\circ$ to 50$^\circ$. Each map covers 120$^\circ$
of longitude, with the centers at longitudes 0$^\circ$ {\it (a)}, 120$^\circ$ {\it (b)},
and 240$^\circ$ {\it (c)}. The upper part of each page shows positive galactic latitudes,
and the lower part shows negative galactic latitudes.
In each panel, the most significant clouds are identified and
labeled using the catalog of Magnani et al. (1997) or, for new clouds 
the `DIRlll$\pm$bb' cloud name. The greyscale ranges from -0.2 (white)
to 5 (black) MJy sr$^{-1}$ at 100 $\mu$m wavelength. Contours are drawn
at 0.3, 1, 2, 3, and 4 MJy sr$^{-1}$ (but the highest levels are not
present in every map). The large black areas are masked, either because of
lack of H~I data, presence of the Magellanic clouds, or presence of extremely
bright and structured infrared emission. \label{fig:exmaps}}

\figcaption[exmappole.ps]{Maps of the 100 $\mu$m infrared excess at 
the galactic poles, in orthographic projections.
Panel {\it (a)} covers the north galactic pole and panel {\it (b)} covers the south
galactic pole. As in the previous figure, the most significant infrared excess 
clouds are labeled. The greyscale range, from white to black, is
-0.2 to 3 MJy sr$^{-1}$. Contours are drawn at 0.1, 1, 0.5, 1, and 2 
MJy sr$^{-1}$. (The contours are at lower values in this figure than in 
Figure 7, in order to bring out fainter features near the poles.)
The large black stripe in panel {\it (b)} is due to a region
where the H~I 21-cm data are poorly sampled. \label{fig:exmappole}}

\figcaption[myst.ps]
{Surface brightness map at 60 $\mu$m of the warm infrared excess cloud
DIR015+54, made from the {\it IRAS Sky Survey
Atlas}. The minimum surface brightness in the map is -1.83 MJy~sr$^{-1}$
(negative because of oversubtraction of zodiacal light), and the contours 
are at -1.35, -1, and -0.65 MJy~sr$^{-1}$. \label{fig:myst}}

\figcaption[extemphis.ps]
{Histograms of color temperatures of MBM clouds (top) and DIRBE clouds (bottom)
obtained from the infrared excess at 100 and 240 $\mu$m. The color temperature
of emission associated with the atomic gas is shown as a dashed line. Both the
MBM clouds, which are known to be molecular, and the new infrared excess clouds
are colder than atomic clouds, most likely because they are optically thick to the
high-energy part of the interstellar radiation field.\label{fig:extemphis}}

\figcaption[figheith.ps]{Comparison of the infrared 
excess (smooth contours) and CO($1\rightarrow 0$) line integral
for the Ursa Major-Ursa Minor-Camelopardalis region. Regions where the CO was observed 
are surrounded
by thick lines and filled with thin diagonal lines. The two maps are in very good 
agreement where
data for both maps exist. The infrared excess predicts, accurately, the location 
and extent of the
molecular clouds. Furthermore, there are no `extra' infrared excess clouds predicted in
 the regions
where CO was observed but not detected.\label{fig:exheith}}

\figcaption[exheithcorr.ps]{Correlation of
the infrared excess, $I_{100}^{ex}$ (MJy~sr$^{-1}$), and CO($1\rightarrow 0$) line integral,
$W($CO$)$ (K~km~s$^{-1}$ for the North Celestial Pole
region. Each point corresponds to an independent $40^\prime\times40^\prime$ pixel.
\label{fig:excorr}}

\begin{deluxetable}{lccccccc}
\footnotesize
\tablecaption{Infrared to H I Correlations for Two Regions (or Cells)\label{tab:irhicell}}
\tablewidth{0pt}
\tablehead{
& \multicolumn{2}{c}{$(l,b)$=$(140^\circ,35^\circ)$} && \multicolumn{2}{c}{$(l,b)$=$(40^\circ,35^\circ)$} \\ 
\cline{2-4}\cline{6-8}
\colhead{Wavelength} & \colhead{Slope} & \colhead{Offset} & \colhead{$\chi_\nu^2$} &&  
                                    \colhead{Slope} & \colhead{Offset} & \colhead{$\chi_\nu^2$}\\
& \colhead{[MJy sr$^{-1}$ / 10$^{20}$ cm$^{-2}$]} & \colhead{[MJy sr$^{-1}$]} & & ~~&
   \colhead{[MJy sr$^{-1}$ / 10$^{20}$ cm$^{-2}$]} & \colhead{[MJy sr$^{-1}$]} &
}
\startdata
100 $\mu$m    & $0.56\pm 0.01$ & $0.63\pm 0.02$ & 1.25 &&  $0.89\pm 0.01$ & $0.33\pm 0.04$ & 1.77\nl
140 $\mu$m    & $1.12\pm 0.06$ & $1.01\pm 0.14$ & 1.06 &&  $1.45\pm 0.06$ & $0.44\pm 0.21$ & 0.91\nl
240 $\mu$m    & $1.02\pm 0.03$ & $0.62\pm 0.08$ & 1.26 &&  $1.07\pm 0.04$ & $0.42\pm 0.12$ & 0.91\nl
\enddata
\end{deluxetable}

\begin{deluxetable}{lccc}
\footnotesize
\tablecaption{Statistics of Slopes and Intercepts\tablenotemark{a} \label{tab:slopehis}}
\tablewidth{0pt}
\tablehead{
\colhead{Wavelength} & \colhead{Slope} & \colhead{Offset} & \colhead{Offset w/o outliers\tablenotemark{b}} \\
& \colhead{[MJy sr$^{-1}$ / 10$^{20}$ cm$^{-2}$]} & \colhead{[MJy sr$^{-1}$]} & \colhead{[MJy sr$^{-1}$]}
}
\startdata
100 $\mu$m    & $0.62\pm 0.08$ & $0.76\pm 0.37$ & $0.71\pm 0.32$ \nl
140 $\mu$m    & $0.95\pm 0.17$ & $1.11\pm 0.63$ & $1.10\pm 0.52$ \nl
240 $\mu$m    & $0.72\pm 0.11$ & $0.87\pm 0.55$ & $0.83\pm 0.41$ \nl
\enddata
\tablenotetext{a}{mean and rms dispersion, from cell to cell}
\tablenotetext{b}{excluding those regions for
which the slope deviates by more than 1.5 times its dispersion from its mean value}
\end{deluxetable}

\begin{deluxetable}{lrrrcc}
\footnotesize
\tablecaption{Infrared Excess from Known High-Latitude Molecular Clouds\label{tab:excatknown}}
\tablewidth{0pt}
\tablehead{
\colhead{Name\tablenotemark{a}} & \colhead{$l$}   & \colhead{$b$}   & \colhead{size} & 
\colhead{$T(240/100)$\tablenotemark{b}} & \colhead{$F_{100}^{ex}$\tablenotemark{b}} \\
\colhead{} & \colhead{} & \colhead{} & \colhead{[deg]} & \colhead{[K]} & \colhead{[kJy]} 
} 
\startdata
\input cloudcat.tex
\enddata
\tablenotetext{a}{Cloud names mostly from the compilation of high-latitude molecular 
clouds (\cite{magnani97}).
MBM: extinction patches on Palomar Observatory Sky Survey (\cite{mbm85}),
KM, DC: extinction patches in European Southern Observatory sky survey (\cite{km86}; \cite{hartley}),
HRK: infrared cirrus patches in IRAS 100 $\mu$m maps (\cite{hrk88})}
\tablenotetext{b}{Values listed as `...' in clouds for which
the \COBE/DIRBE data indicates strong, unresolved structure (consistent with
a bright point source.)}
\end{deluxetable}

\begin{deluxetable}{lrrrcc}
\footnotesize
\tablecaption{Catalog of Unidentified Infrared Excess Clouds\label{tab:excatnew}}
\tablewidth{0pt}
\tablehead{
\colhead{Name} & \colhead{$l$}   & \colhead{$b$}   & \colhead{size} & 
\colhead{$T(240/100)$}  & \colhead{$F_{100}^{ex}$} \\
\colhead{} & \colhead{} & \colhead{} & \colhead{[deg]} & \colhead{[K]} & \colhead{[kJy]} 
} 
\startdata
\input cloudcat2.tex
\enddata
\end{deluxetable}

\begin{deluxetable}{lccccc}
\footnotesize
\tablecaption{Multiple Linear Regression between Infrared, H~I, and CO 
Maps\tablenotemark{a} \label{tab:excorr}}
\tablewidth{0pt}
\tablehead{
\colhead{Wavelength} & \colhead{$a=I_\nu/N({\rm H~I})$} & 
\colhead{$b=I_\nu/W({\rm CO})$} & \colhead{$\chi_\nu^2$} & 
\multicolumn{2}{c}{$X=N({\rm H}_2)/W({\rm CO})$} \\ 
 & \colhead{[MJy sr$^{-1}$ / 10$^{20}$ cm$^{-2}$]} & 
 \colhead{[MJy sr$^{-1}$ / (K~km~s$^{-1}$)]} & 
     & \multicolumn{2}{c}{[10$^{20}$ cm$^{-2}$/(K~km~s$^{-1}$)]}  \\ \cline{5-6} 
&&&& \colhead{na\"{\i}ve} & \colhead{true}
}
\startdata
100 $\mu$m    & $0.48\pm 0.02$ & $0.34\pm 0.03$ & 1.1 & $0.35\pm 0.05$ & $1.3\pm 0.3$ \nl
140 $\mu$m    & $1.06\pm 0.08$ & $1.24\pm 0.12$ & 0.8 & $0.58\pm 0.11$ & $1.5\pm 0.4$ \nl
240 $\mu$m    & $0.94\pm 0.06$ & $1.33\pm 0.10$ & 1.2 & $0.71\pm 0.11$ & $1.3\pm 0.2$ \nl
\enddata
\tablenotetext{a}{For the Ursa Major/Ursa Minor/Camelopardalis region}
\end{deluxetable}

\end{document}

%% file: cloudcat.tex
DC001.4-21.6 &   1.4 & -21.6 & 0.6 & 16.2 &   1.31 \nl
R Cor Bor      &   4.0 & -25.5 & 2.0 & 17.1 &   ... \nl
LDN 134         &   5.0 &  36.0 & 3.0 & 14.8 &   ... \nl
MBM 45        &   9.8 & -28.0 & 1.0 & 14.2 &   0.72 \nl
MBM 39        &  11.4 &  36.2 & 1.9 & 16.6 &   1.56 \nl
MBM 40        &  37.6 &  44.7 & 2.2 & 15.9 &   0.48 \nl
MBM 46/48     &  40.3 & -35.3 & 2.0 & 16.9 &   2.12 \nl
HD 210121     &  57.5 & -44.0 & 4.0 & 16.3 &   3.57 \nl
G61-34       &  61.3 & -34.0 & 1.3 & 15.2 &   1.16 \nl
MBM 49        &  64.7 & -26.8 & 1.0 & .... & ... \nl
MBM 50        &  70.0 & -30.7 & 5.5 & 21.0 &   1.85 \nl
G72+25       &  71.0 &  25.5 & 1.0 & 17.7 &   ... \nl
MBM 51/52     &  74.0 & -51.3 & 4.0 & 17.0 &   0.66 \nl
MBM 55        &  89.2 & -40.9 & 3.4 & 14.6 &   1.77 \nl
MBM 41/44     &  90.0 &  38.9 & 1.8 & 15.7 &   ... \nl
MBM 53        &  93.0 & -32.2 & 4.4 & 19.0 &   ... \nl
MBM 54        &  93.0 & -37.5 & 4.3 & 15.9 &   ... \nl
HRK 94.8+38   &  94.8 &  37.5 & 1.2 & .... & ... \nl
MBM 56        & 102.7 & -28.1 & 2.5 & 17.2 &   1.24 \nl
MBM 1         & 110.2 & -41.2 & 0.5 & 17.8 &   0.57 \nl
MBM 2         & 117.7 & -52.6 & 1.9 & 19.0 &   1.28 \nl
HSVMT 1       & 119.1 &  28.5 & 1.8 & 15.9 &   0.61 \nl
Polaris      & 125.0 &  28.0 & 5.0 & 15.1 &   0.79 \nl
HSVMT 12/14   & 125.5 &  32.5 & 1.0 & 16.0 &   0.33 \nl
MBM 3         & 130.0 & -46.8 & 2.5 & 15.4 &   1.21 \nl
MBM 4         & 134.2 & -45.0 & 2.0 & 14.7 &   0.63 \nl
HRK 135+51    & 135.6 &  51.3 & 1.5 & 16.2 &   0.62 \nl
HRK 135-69    & 136.4 & -68.3 & 1.8 & 17.9 &   2.32 \nl
HRK 135+54    & 136.5 &  54.8 & 1.1 & 17.5 &   2.86 \nl
HRK 140+48    & 140.5 &  48.0 & 0.7 & 24.4 &   0.74 \nl
MBM 27/29     & 142.0 &  35.0 & 2.2 & 15.6 &   ... \nl
MBM 30        & 142.2 &  38.2 & 3.4 & 15.2 &   0.50 \nl
Camelopardalis & 143.5 &  24.0 & 2.9 & 16.0 &   0.61 \nl
MBM 5         & 145.0 & -49.9 & 3.8 & 13.8 &   0.60 \nl
MBM 6         & 147.0 & -39.0 & 1.9 & 15.0 &   1.07 \nl
MBM 31/32     & 147.0 &  40.0 & 4.2 & 15.9 &   ... \nl
MBM 7-8       & 150.9 & -38.3 & 1.5 & 14.8 &   2.87 \nl
HSVMT 27      & 153.6 &  36.9 & 2.8 & 15.7 &   2.20 \nl
G155-40      & 154.7 & -39.8 & 1.4 & 15.2 &   0.91 \nl
MBM 26        & 156.4 &  32.6 & 2.0 & 18.1 &   ... \nl
MBM 11/14     & 159.0 & -33.6 & 3.5 & 15.3 &   0.62 \nl
MBM 17        & 165.7 & -25.6 & 0.5 & .... & ... \nl
MBM 16        & 171.2 & -37.7 & 0.5 & 15.5 &   0.77 \nl
MBM 23/24     & 172.0 &  26.9 & 0.8 & 15.5 &   0.84 \nl
MBM 25        & 173.6 &  31.2 & 2.0 & 15.9 &   ... \nl
MBM 19        & 186.0 & -29.9 & 0.5 & .... & ... \nl
MBM 18        & 189.1 & -36.0 & 0.5 & .... & ... \nl
MBM 15        & 191.7 & -51.3 & 3.5 & 17.7 &   ... \nl
HRK 192-67    & 192.0 & -67.5 & 1.4 & 18.5 &   2.29 \nl
CB 28         & 204.0 & -25.2 & 1.5 & 14.9 &   0.28 \nl
MBM 21-22     & 208.3 & -28.0 & 2.0 & 19.6 &   0.53 \nl
G211+63      & 210.8 &  63.1 & 2.0 & 18.4 &   ... \nl
MBM 20        & 210.9 & -36.5 & 2.5 & 15.4 &   2.27 \nl
HRK 225-66    & 227.0 & -66.1 & 2.5 & 15.6 &   2.69 \nl
HRK 228-29    & 229.0 & -28.5 & 1.7 & 16.4 &   ... \nl
HRK 236+39    & 236.0 &  38.0 & 4.0 & 16.5 &   0.22 \nl
KM 273+29     & 272.5 &  29.7 & 2.0 & 17.9 &   0.65 \nl
KM 293-31     & 291.0 & -31.0 & 2.2 & .... & ... \nl
KM 300-24     & 301.1 & -24.5 & 2.0 & .... & ... \nl
MBM 33        & 359.1 &  36.7 & 1.2 & 15.7 &   ... \nl

%% file: cloudcat2.tex
   DIR002+31 &   1.6 & -30.8 & 1.2 & 18.4 &   0.89 \nl
   DIR009+30 &   9.1 & -30.1 & 2.3 & 16.5 &   2.50 \nl
   DIR013+40 &  12.8 &  40.0 & 1.5 & 17.1 &   1.31 \nl
   DIR018+37 &  17.8 & -36.5 & 2.5 & 12.2 &   0.04 \nl
   DIR020+45 &  19.5 & -44.5 & 1.7 & 18.9 &   0.69 \nl
   DIR021+52 &  21.0 &  52.0 & 1.5 & 17.5 &   0.29 \nl
   DIR025+35 &  25.0 &  35.0 & 3.5 & 21.8 &   2.96 \nl
   DIR027+31 &  26.5 & -30.5 & 1.0 & 15.3 &   0.52 \nl
   DIR028+54 &  28.3 &  53.7 & 1.0 & 17.3 &   0.29 \nl
   DIR029+25 &  28.5 &  25.0 & 2.6 & 16.6 &   2.15 \nl
   DIR029+30 &  28.5 &  30.0 & 1.1 & 16.7 &   0.72 \nl
   DIR034+26 &  34.0 &  26.0 & 3.0 & 18.6 &   3.21 \nl
   DIR046+37 &  45.5 & -36.5 & 1.6 & 16.1 &   1.24 \nl
   DIR046+33 &  45.7 & -33.0 & 2.2 & 22.4 &   0.72 \nl
   DIR048+25 &  47.7 &  24.5 & 2.1 & 16.9 &   1.49 \nl
   DIR048+38 &  48.0 &  38.0 & 2.5 & 18.0 &   1.56 \nl
   DIR060+26 &  59.7 & -26.0 & 1.7 & 22.3 &   1.07 \nl
   DIR061+22 &  61.0 &  22.0 & 4.0 & 26.1 &   3.57 \nl
   DIR070+23 &  69.7 &  22.5 & 1.2 & 17.2 &   0.63 \nl
   DIR071+43 &  71.0 & -42.5 & 3.0 & 15.5 &   3.14 \nl
   DIR072+34 &  71.5 & -34.0 & 1.4 & 15.9 &   0.43 \nl
   DIR077+37 &  76.5 & -37.0 & 3.3 & 18.5 &   1.75 \nl
   DIR081+39 &  81.3 &  38.5 & 2.5 & 16.2 &   0.80 \nl
   DIR087+29 &  86.5 &  28.5 & 3.5 & 23.1 &   1.07 \nl
   DIR096+23 &  96.2 &  23.0 & 2.0 & 19.3 &   1.17 \nl
   DIR098+44 &  98.0 & -44.0 & 2.5 & 17.4 &   0.97 \nl
   DIR105-31 & 105.0 & -31.0 & 1.8 & 15.3 &   1.16 \nl
   DIR105-38 & 105.0 & -38.0 & 2.7 & 15.7 &   2.48 \nl
   DIR108-27 & 108.0 &  27.0 & 2.7 & 18.1 &   2.12 \nl
   DIR117-44 & 116.5 & -44.0 & 2.0 & 18.0 &   0.70 \nl
   DIR120-28 & 120.0 & -28.0 & 2.4 & 17.7 &   1.28 \nl
   DIR121-45 & 120.9 & -45.0 & 1.8 & 15.5 &   0.57 \nl
   DIR126-37 & 126.0 &  37.0 & 0.9 & 16.7 &   0.30 \nl
   DIR132-30 & 131.5 & -29.5 & 1.9 & 18.4 &   0.61 \nl
   DIR134-36 & 134.0 & -35.5 & 1.5 & 16.3 &   0.62 \nl
   DIR135-38 & 134.5 &  38.0 & 1.4 & 23.4 &   0.28 \nl
   DIR135-41 & 134.7 &  40.6 & 1.4 & 17.0 &   0.34 \nl
   DIR140-45 & 140.0 & -44.5 & 2.0 & 19.8 &   0.63 \nl
   DIR150-29 & 150.2 & -28.5 & 1.7 & 16.2 &   0.89 \nl
   DIR152-47 & 152.0 & -46.5 & 3.0 & 16.2 &   2.74 \nl
   DIR164-44 & 163.5 & -44.0 & 1.6 & 17.0 &   0.61 \nl
   DIR172-42 & 172.0 & -41.5 & 2.3 & 15.6 &   2.87 \nl
   DIR177-33 & 177.2 &  33.0 & 1.5 & 16.3 &   0.56 \nl
   DIR179-49 & 179.0 & -49.0 & 3.5 & 20.4 &   1.68 \nl
   DIR184-26 & 183.5 &  26.0 & 1.8 & 17.8 &   0.66 \nl
   DIR187-43 & 187.0 & -42.5 & 1.7 & 20.4 &   0.58 \nl
   DIR196-24 & 196.0 &  24.2 & 1.0 & 17.2 &   0.46 \nl
   DIR198-32 & 198.0 &  32.0 & 1.5 & 15.8 &   0.61 \nl
   DIR201-24 & 200.5 &  23.6 & 1.3 & 19.9 &   0.39 \nl
   DIR203-32 & 202.5 & -31.5 & 2.0 & 14.9 &   0.33 \nl
   DIR204-37 & 203.5 & -36.5 & 1.8 & 15.6 &   0.79 \nl
   DIR216+27 & 216.3 &  26.5 & 2.2 & 16.2 &   0.34 \nl
   DIR223+37 & 223.0 & -37.0 & 2.7 & 22.3 &   1.21 \nl
   DIR234+37 & 234.0 & -37.0 & 1.1 & 21.2 &   0.65 \nl
   DIR237+44 & 237.0 & -44.0 & 1.2 & 19.7 &   0.72 \nl
   DIR239-25 & 239.0 & -25.0 & 4.0 & 20.5 &   3.34 \nl
   DIR245+35 & 245.0 &  34.8 & 2.1 & 17.1 &   0.88 \nl
   DIR257+34 & 256.5 &  34.0 & 1.5 & 16.2 &   0.76 \nl
   DIR265+31 & 265.0 & -30.9 & 3.0 & 16.2 &   1.96 \nl
   DIR274+46 & 274.0 & -45.8 & 1.4 & 18.1 &   0.63 \nl
   DIR276+33 & 275.7 &  33.4 & 1.2 & 18.4 &   0.52 \nl
   DIR280+55 & 280.0 & -55.0 & 3.1 & 17.0 &   0.91 \nl
   DIR281-40 & 280.5 &  39.5 & 2.0 & 23.6 &   0.48 \nl
   DIR282+41 & 282.0 & -41.0 & 4.0 & 17.6 &   3.42 \nl
   DIR288+32 & 287.5 &  32.0 & 3.0 & 18.5 &   2.12 \nl
   DIR289-53 & 289.0 & -53.0 & 3.5 & 16.2 &   0.95 \nl
   DIR290-62 & 290.0 & -62.0 & 4.5 & 18.3 &   2.20 \nl
   DIR292-37 & 292.0 & -37.0 & 3.0 & 17.7 &   2.71 \nl
   DIR310+39 & 310.0 &  39.0 & 1.9 & 16.8 &   0.99 \nl
   DIR313-34 & 313.0 & -34.0 & 1.5 & 18.1 &   0.62 \nl
   DIR314-47 & 314.0 & -46.5 & 4.4 & 19.8 &   2.76 \nl
   DIR316+39 & 316.1 &  38.5 & 1.4 & 18.6 &   0.61 \nl
   DIR321+36 & 321.0 & -36.0 & 2.5 & 19.2 &   2.32 \nl
   DIR327+30 & 327.0 & -30.0 & 3.0 & 17.6 &   2.86 \nl
   DIR331-34 & 331.0 & -33.5 & 2.1 & 21.4 &   0.69 \nl
   DIR333+36 & 333.0 & -36.0 & 1.9 & 17.0 &   0.74 \nl
   DIR335-40 & 334.5 & -40.0 & 1.3 & 19.1 &   0.71 \nl
   DIR340-43 & 339.5 & -43.3 & 1.4 & 16.0 &   0.50 \nl
   DIR349-46 & 348.5 & -46.0 & 1.9 & 20.1 &   0.93 \nl
   DIR354-37 & 354.0 &  36.5 & 1.6 & 17.8 &   0.72 \nl
   DIR357-29 & 357.0 & -29.0 & 1.3 & 15.7 &   0.60 \nl

%% file: ms.bbl
\begin{thebibliography}{}

\bibitem[Arendt et al. 1998]{arendt98} Arendt, R. G. et al. 1998, ApJ, submitted

\bibitem[Benjamin et al. 1996]{benjamin96} Benjamin, R. A., Venn, K. A., Hiltgen, D. D., 
\& Sneden, C. 1996, ApJ, 464, 836

\bibitem[Blitz et al. 1984]{blitz84} Blitz, L., Magnani, L., \& Mundy, L. 1984, ApJL, 282, L9


\bibitem[Boggess et al. 1992]{boggess92} Boggess, N. W., Mather, J. C., 
Weiss, R., Bennett, C. L., Cheng, E. S.,
Dwek, E., Gulkis, S., Hauser, M. G., Janssen, M. A., Kelsall, T., Meyer,
S. S., Moseley, S. H., Murdock, T. L., Shafer, R. A., Silverberg, R. F.,
Smoot, G. F., Wilkinson, D. T. and E. L. Wright 1992, ApJ, 397, 420

\bibitem[Bohlin et al. 1978]{bohlin} Bohlin, R. C., Savage, B. D., \& Drake, J. F. 1978, ApJ, 224, 132

\bibitem[Boulanger and P\'erault 1988]{bp88} Boulanger, F., \& P\'erault, M. 1988, ApJ, 330, 964

\bibitem[Boulanger et al. 1998]{boulcham} Boulanger, F., Bronfman, L., Dame, T. M., \& 
Thaddeus, P. 1998, A\&A, 332, 273

\bibitem[Boulanger et al. 1996]{boulanger96} Boulanger, F., Abergel, A., Bernard, J. -P.,
 Burton, W. B., D\'esert, F.-X., Hartmann, D., Lagache, G., Puget, J.-L. 1996, A\& A, 312, 256
 
\bibitem[Burstein and Heiles 1978]{burstein} Burstein, D., \& Heiles, C. 1978, ApJ, 225, 40

\bibitem[Cleary et al. 1979]{cleary} Cleary, M. N., Heiles, C., \& Haslam, C. G. T. 1979, A\& AS, 36, 95

\bibitem[D\'esert et al. 1988]{dbb88} D\'esert, F. X., Bazell, D., \& Boulanger, F. 1988, ApJ, 334, 815

\bibitem[D\'esert et al. 1990]{dbp90} D\'esert, F. X., Boulanger, F., \& Puget, J.-L. 1990, A\& A, 327, 215

\bibitem[de Vries et al. 1987]{vht87} de Vries, H. W., Thaddeus, P., \& Heithausen, A. 1987, ApJ, 319, 723

\bibitem[Digel et al. 1996]{digel96} Digel, S. W., Grenier, I. A., Heithausen, A., Hunter, S. D.,
 \& Thaddeus, P. 1996, ApJ, 463, 609
 
\bibitem[Draine and Lee 1984]{dl84} Draine, B. T., \& Lee, H. M. 1985, ApJ, 285, 89

\bibitem[Dwek et al. 1998]{Dwek98} Dwek, E., Arendt, R. G., Hauser, M. G., Fixsen, D., Kelsall, T.,
Leisawitz, D., Pei, Y. C., Wright, E. L., Mather, J. C., Moseley, S. H., Odegard, N.,
Shafer, R., Silverberg, R. F., \& Weiland, J. L. 1998, ApJ, submitted

\bibitem[ESA 1997]{esa97} ESA 1997, {\it The Hipparcos and Tycho Catalogues}, ESA SP-1200

\bibitem[Gir et al. 1994]{gir94} Gir, B.-Y., Blitz, L., \& Magnani, L. 1994, ApJ, 434, 162

\bibitem[Hartley et al. 1986]{hartley} Hartley, M., Tritton, S. B., Manchester, R. N., Smith, R. M.,
\& Goss, W. M. 1986, A\& AS, 63, 27

\bibitem[Hartmann and Burton 1997]{hartmann97} Hartmann, D., \& Burton, W. B. 1997,
{\it Atlas of Galactic Neutral Hydrogen} (Cambridge University Press: Cambridge)

\bibitem[Hartmann et al. 1996]{hartstray} Hartmann, D., Kalberla, P. M. W., Burton, W. B., \&
Mebold, U. 1996, A\& AS, 312, 256

\bibitem[Hartmann et al. 1998]{hartmag} Hartmann, D., Magnani, L., \& Thaddeus, P. 1998,
  ApJ, 492, 205
  
\bibitem[Hauser et al. 1997]{dexpsup} Hauser, M. G., Kelsall, T., Leisawitz, D., \&
Weiland, J. L., eds., {\it COBE Diffuse Infrared Background Experiment Explanatory
Supplement}, version 2.1, COBE reference publication \#97-A (Greenbelt: NASA/GSFC),
available in electronic form from the National Space Science Data Center

\bibitem[Hauser et al. 1998]{hauser98} Hauser, M. G., Arendt. R. G., Kelsall, T.,
Dwek, E., Odegard, N., Weiland, J. L., Freudenreich, H. T., Reach, W. T., Silverberg,
R. F., Moseley, S. H., Pei, Y. C., Mather, J. C., Smoot, G. F., Wilkinson, D. T.,
\& Wright, E. L. 1998, ApJ, submitted

\bibitem[Heiles and Habing 1974]{hh} Heiles, C., \& Habing, H. J. 1974, A\& AS, 14, 1

\bibitem[Heiles et al. 1988]{hrk88} Heiles, C., Reach, W. T., \& Koo, B.-C. 1988, ApJ, 332, 313 (HRK)

\bibitem[Heithausen \& Thaddeus 1990]{heith90} Heithausen, A., \& Thaddeus, P. 1990, ApJL, 353, L49

\bibitem[Heithausen et al. 1993]{heith93} Heithausen, A., Stacy, J. G., de Vries, H. W., 
Mebold, U., \& Thaddeus, P. 1993, A\& A, 268, 265

\bibitem[{\it IRAS} 1988]{psc} {\it IRAS} Point Source Catalog, Version 2, 1988, Joint {\it IRAS}
Science Working Group (Washington, DC: GPO)

\bibitem[Israel et al. 1996]{israel96} Israel, F. P., Bontekoe, T. R., \& Kester, D. J. M. 1996, A\& A,308, 723

\bibitem[Kawara et al. 1997]{kawara} Kawara, K., Taniguchi, Y., Sato, Y., Okuda, H., Matsumoto, T.,
Sofue, Y., Wakamatsu, K., Matsuhara, H., Hasegawa, T., Chambers, K. C., Cowie, L. L.,
Joseph, R. D., Sanders, D. B., \& Wynn-Williams, C. G. 1997, in {\it Diffuse Infrared
Radiation and the IRTS}, eds. H. Okuda, T. Matsumoto, \& T. Rollig (San
Francisco: ASP), p. 386

\bibitem[Kelsall et al. 1998]{kelsall98} Kelsall, T., Weiland, J. L., Franz, B. A.,
Reach, W. T., Arendt, R. G., Dwek, E., Freudenreich, H. T., Hauser, M. G., 
Moseley, S. H., Odegard, N. P., \& Silverberg, R. F. 1998, ApJ, submitted

\bibitem[Keto and Myers 1986]{km86} Keto, E. R., \& Myers, P. C. 1986, ApJ, 304, 466

\bibitem[Kulkarni and Heiles 1988]{kulkarni88} Kulkarni, S. R., \& Heiles, C. 1988, in
{\it Galactic and Extragalactic Radio Astronomy: Second Edition}, ed. G. L. Verschuur and
K. I. Kellermann (Berlin: Springer-Verlag), 95

\bibitem[Lagache et al. 1998]{lagache98} Lagache, G., Abergel, A., Boulanger, F., \& Puget, J.-L. 1998,
A\& A, in press

\bibitem[Laureijs et al. 1996]{Laureijs96} Laureijs, R. J., Haikala, L., Burgdorf, M., Clark, F. O.,
Liljestr\"om, T., Mattila, K., \& Prusti, T. 1996, A\& A 315, L317

\bibitem[Low et al. 1984]{low} Low, F. J., et al. 1984, ApJL, 278, L19

\bibitem[Luhman and Jaffe 1996]{luhman96} Luhman, M. L., \& Jaffe, D. 1996, ApJ, 463, 191

\bibitem[Magnani et al. 1985]{mbm85} Magnani, L., Blitz, L., \& Mundy, L. 1985, ApJ, 295, 402 (MBM)

\bibitem[Magnani et al. 1988]{mbw88} Magnani, L., Blitz, L., \& Wouterloot, J. G. A. 1988, ApJ, 295, 402

\bibitem[Magnani et al. 1996]{magnani97} Magnani, L., Hartmann, D., \& Speck, B. G. 1996, ApJS, 106, 447

\bibitem[Magnani et al. 1986]{magnani86} Magnani, L., Lada, E. A., \& Blitz, L. 1986, ApJ, 301, 395

\bibitem[Magnani and Onello 1995]{magnani95}Magnani, L., \& Onello, J. S. 1995, ApJ, 443, 169

\bibitem[McCammon and Sanders 1990]{mccammon} McCammon, D., \& Sanders, W. T. 1990, ARA\& A, 28, 657

\bibitem[Moshir et al. 1989]{fsc} Moshir, M. et al. 1992, {\it Explanatory Supplement to the
IRAS Faint Source Survey}, Version 2, JPL, Pasadena

\bibitem[Neugebauer et al. 1984]{neugebauer} Neugebauer, G., et al. 1984, ApJL, 278, L1

\bibitem[Odenwald et al. 1997]{odenwald97} Odenwald, S., Newmark, J., \& Smoot, G. 1997,
preprint astro-ph/9610238

\bibitem[Puget et al. 1996]{puget96} Puget, J.-L., Abergel, A., Bernard, J. -P., Boulanger, F.
 Burton, W. B., D\'esert, F.-X., Hartmann 1996, A\& A, 308, 5
 
\bibitem[Puget et al. 1998]{pugfirback} Puget, J.-L., Lagache, G., Clements, D., Reach, W. T.,
Aussel, H., Bouchet, F., Cesarsky, C., D\'esert, F. X., Elbaz, D., Franceschini, A., \&
Guiderdoni, B. 1998, in preparation

\bibitem[Reach et al. 1993]{cps}Reach, W. T., Heiles, C., \& Koo, B.-C. 1993, ApJ, 412, 127

\bibitem[Reach et al. 1994]{rkh94}Reach, W. T., Koo, B.-C., \& Heiles, C. 1994, ApJ, 429, 672

\bibitem[Reach et al. 1996]{reach96} Reach, W. T., Franz, B. A., 
Kelsall, T., and J. L. Weiland 1996,
in {\it Unveiling the Cosmic Infrared Background}, ed. E. Dwek
(New York: AIP), 37

\bibitem[Reach et al. 1994]{reachbaas} Reach, W. T., Wall, W. F., Hauser, M. G.,
Sodroski, T. J., Odegard, N., \& Weiland, J. L. 1994, BAAS, 185, 57.09

\bibitem[Reynolds 85]{reynoldsspica} Reynolds, R. J. 1985, AJ, 90, 92

\bibitem[Sargent et al. 1988]{sargent} Sargent, W. L. W., Steidel, C. C., \& Bocksenberg, A. 1988,
ApJS, 68, 439

\bibitem[Savage et al. 1977]{savage77} Savage, B. D., Bohlin, R. C., Drake, J. F., \& Budich, W.
1977, ApJ, 216, 291

\bibitem[Schlegel et al. 1998]{schlegel97} Schlegel, D. Finkbeiner, D. P., \& Davis, M. 1998, ApJ, in press

\bibitem[Shafer et al. 1998]{shafer98} Shafer, R., et al. 1998, in preparation 

\bibitem[Snowden et al. 1993]{snowden93} Snowden, S. L., McCammon, D., \& Verter, F. 1995, ApJL, 309, L21

\bibitem[Sodroski et al. 1995]{sodroski95} Sodroski, T. J., Odegard, N., Dwek, E., 
Hauser, M. G., Freedman, J., Kelsall, T., Wall, W. F.,
Berriman, G. B., Odenwald, S. F., Bennett, C., Reach, W. T., \& Weiland, J. L., 1995, ApJ, 452, 262

\bibitem[Stark 1995]{stark95} Stark, R. 1995, A\& A, 301, 873

\bibitem[Strong et al. 1994]{strong94} Strong, A. W., Bennett, K., Bloemen, H., Diehl, R., Hermsen, W., 
Morris, D., Schoenfelder, V., Stacy, J. G., de Vries, C., Varendorff, M., Winkler, C.,
                        Youssefi, G. 1994, A\&A, 292, 82

\bibitem[Strong et al. 1988]{strong88} Strong, A. W., et al. 1988, A\&A, 207, 1

\bibitem[van Buren and McCray 1988]{dvb88} van Buren, D., \& McCray, R. 1988, ApJL, 336, L67

\bibitem[Wakker and Boulanger 1986]{boulwakk} Wakker, B. P., \& Boulanger, F. 1986, A\& A, 170, 84

\bibitem[Wall et al. 1996]{Wall} Wall, W. F., Reach, W. T., Hauser, M. G., Arendt, R. G.,
 Weiland, J. L., Berriman, G. B., Bennett, C. L., Dwek, E., Leisawitz, D., Mitra, P. M.,
 Odenwald, S. F., Sodroski, T. J., \& Toller, G. N. 1996, ApJ, 456, 566

\bibitem[Wang and Yu 1995]{wang95} Wang, Q. D., \& Yu, K. C. 1995, AJ 109, 698

\bibitem[Zagury 1997]{zagury} Zagury, F. 1997, Ph. D. thesis, Universit\'e Paris-Sud (Orsay, France)

\end{thebibliography}
